\documentstyle[12pt,titlepage,epsf]{article}
\newcommand{\nin}{\noindent}
\newcommand{\be}{\begin{equation}}

\newcommand{\ee}{\end{equation}}
\newcommand{\bea}{\begin{eqnarray}}
\newcommand{\eea}{\end{eqnarray}}
\newcommand{\br}{\hskip .25cm/\hskip -.25cm}
\newcommand{\hf}{\frac{1}{2}}
\newcommand{\nonu}{\nonumber\\}
\newcommand{\la}{\hskip .25cm^\leftarrow\hskip -.25cm}
\newcommand{\dg}{^\dagger}
\newcommand{\ol}{\overline}
\def\gsim{\mathrel{\rlap {\raise.5ex\hbox{$ > $}}
{\lower.5ex\hbox{$\sim$}}}}
\def\lsim{\mathrel{\rlap {\raise.5ex\hbox{$ < $}}
{\lower.5ex\hbox{$\sim$}}}}

\def\gappeq{\mathrel{\rlap {\raise.5ex\hbox{$>$}}
{\lower.5ex\hbox{$\sim$}}}}

\def\lappeq{\mathrel{\rlap{\raise.5ex\hbox{$<$}}
{\lower.5ex\hbox{$\sim$}}}}

\begin{document}
\begin{titlepage}

\begin{flushright}
hep-ph/0107223
\end{flushright}

\vspace{0.1in}
\begin{centering}

{\Large {\bf  Spatially Anisotropic Four-Dimensional
Gauge Interactions, Planar Fermions and
Magnetic Catalysis}}
 \\
\vspace{0.2in}
{\bf J. Alexandre},
{\bf K. Farakos}, {\bf G. Koutsoumbas} \\
\vspace{0.1in}
National Technical University of Athens, Zografou Campus, Athens GR157 80, Greece \\
\vspace{0.1in}
and  \\
{\bf N. E. Mavromatos } \\
\vspace{0.1in}
Department of Physics, Theoretical Physics, King's College London \\
Strand, London WC2R 2LS, U.K. \\

\vspace{0.4in}
{\bf Abstract}

\end{centering}

{\small
We consider magnetic catalysis in a
field-theoretic system of
(3+1)-dimensional 
Dirac fermions with anisotropic kinetic term.
By placing the system in a
strong external magnetic field,
we examine magnetically-induced fermion mass generation.
When the coupling anisotropy
is strong, in which case the fermions effectively 
localize on the plane, 
we find a significant enhancement of the induced mass gap
compared to the isotropic four-dimensional
case of quantum electrodynamics. 
As expected on purely dimensional grounds, the
mass and critical temperature
scale with the square root of the magnetic field.
This phenomenon might be related to recent experimental findings
on magnetically-induced gaps at the nodes of $d$-wave superconducting
gaps in high-temperature cuprates.}

\vspace{0.90in}

\end{titlepage}

\section{Introduction}

The phenomenon of magnetic catalysis,
first suggested in \cite{gusynin}, and
subsequently developed also in \cite{lee}, namely the
dynamical generation of a fermion mass gap
in the presence of external magnetic fields,
has wide applications, ranging from
particle physics and physics of the Early Universe~\cite{applic}
to condensed matter~\cite{farakos,fkm,others}.

In the case of four-dimensional Abelian $U(1)$ gauge-Dirac-fermion
models
the magnetic catalysis phenomenon,
for an external magnetic field $B$, is known~\cite{gus} to yield a
relatively small mass gap $m_{4d}=m_{4d}(B)$ 
(for standard low-energy electromagnetism, with fine structure constant
of order 1/137)
and  an
associated critical temperature of the same order
$T_s \simeq m_{4d}$.
Such analyses
pertain to Abelian gauge interactions with spatially isotropic couplings.
It is the purpose of the present article to consider
{\it spatially anisotropic gauge interaction},
in such a way that the gauge
coupling on a spatial plane,
defined by, say, the $x,y$ directions of a three-dimensional space,
is much stronger than the coupling along the $z$-direction.

Physical motivation for such studies is provided by recent
experimental findings in the physics of high-temperature layered
superconductors~\cite{krishana}. According to such experiments,
one looks at the thermal conductivity properties of a sample of
high-temperature superconductors in the superconducting phase, in
the presence of strong external magnetic fields (with intensities
up to a few Tesla).  The high-temperature superconducting samples
are known to be strongly type II, and so the magnetic field lines
penetrate the material significantly. Moreover, such materials are
known to be $d$-wave superconductors with a layered structure of
Cu-O planes, characterized by nodes in their superconducting
excitation gaps. The experiments of \cite{krishana} have
demonstrated that below a given temperature, which is smaller than
the critical temperature of the superconductor, and which scales
as the square root of the applied magnetic field, there are
plateaux in the thermal conductivity  diagrams. The plateaux
indicate the opening of gaps at the nodes below this new
`critical' temperature (not to be confused with the
superconducting/normal phase critical temperature) .

In ref. \cite{farakos} we suggested that this phenomenon might be
a straightforward application of the magnetic catalysis 
phenomenon~\cite{gusynin,lee,miransky} to the three-dimensional effective 
problem of
physics near the nodes. Indeed, linearizing the excitations about
such nodes one obtains, upon the assumption of spin-charge
separation~\cite{anderson}, a relativistic electrically-charged
fermion (Dirac) system, coupled to statistical gauge fields in the
presence of external electromagnetic fields. The strong external
magnetic field induces the opening of a nodal holon gap, which
scales with the magnetic field $B$. The experiments of
\cite{krishana} have shown a square root scaling with $B$. In the
models of \cite{farakos,fm}, the statistical gauge fields
represent effective spin-spin antiferromagnetic interactions,
which are believed by many to be relevant for the physics of the
high-temperature superconductivity. In ref. \cite{farakos}, and in
all the subsequent works~\cite{fkm,others}, where various
interactions, including four-fermi, among the holons have been
considered, the model systems have been assumed to live
exclusively in $(2+1)$-dimensions, ignoring completely any
four-dimensional physics effects.

This may not be physically correct, especially from the point of
view of the gauge and electromagnetic interactions,
which are known to be fully four-dimensional~\footnote{For instance,
interlayer couplings
via magnetic spin-spin interactions are known to exist
in the planar high-temperature superconducting materials,
and, of course,
the electromagnetic interactions are fully four dimensional.}.
Moreover, the superconducting gaps are actually four-dimensional,
although strongly anisotropic, i.e. with dominant components
along the Cu-O planes, but with suppressed, however non-vanishing, 
components along the perpendicular (interplanar) direction.  

It is the purpose of this article therefore to consider
such an anisotropic four-dimensional 
situation and study the consequences for
magnetically induced dynamical mass generation for the fermionic excitations.
From the condensed-matter point of view, such excitations 
may be the (continuum limit of) holons,
carrying the electric charge only, but no actual spin (in the 
microscopic sense).
In ref. \cite{farakos} 
the holons were assumed purely three-dimensional (planar),
as a result of the localization of their wavefunctions
on the superconducting planes. This is
a basic feature assumed to
characterize the microscopic physics behind spin-charge
separation~\cite{anderson}.
In the present article the fermions will be assumed 
four-dimensional, but with anisotropic kinetic terms,
which allows for interplanar hoping. 
In the strong anisotropic case one should recove the
three-dimensional case. 
In this sense one should have the extension 
of the concept of spin-charge separation to four dimensions,
in the strongly anisotropic case. This, of course, 
does not apply to the concept of fractional statistics~\cite{fm},
which is only an exclusive feature of the three-dimensional case.
However, there may be a different scenario, in which 
the fermions discussed here are viewed as related
to real nodal electron excitations (carrying both spin and charge),
capable of interlayer hopping. 
It is because of this latter interpretation 
that, when we discuss in section 5 
the non-relativistic situation of  relevance to condensed matter,
we shall ignore any statistical gauge interactions among the fermions,
keeping only the electromagnetic interactions.

In case one assumes the existence of four-dimensional holons,
by extending the concept of spin-charge separation to four-dimensions,
allowing, though, for weak interlayer hopping of holons,  
one might encounter a situation relevant to the fully relativistic case 
discussed in the next section, 
in which the gauge fluctuations represent statistical gauge fields.
In such a case one may even 
encounter 
non-abelian gauge interactions. 
Indeed, in the physical models of \cite{fm}, the statistical gauge fields
are non-Abelian, of $SU(2)$ type, due to an underlying
`particle-hole symmetric' formulation of the spin-charge
separation ans\"atze. Such interactions are not responsible for
the opening of holon mass gaps, because of the fermion spectrum in
the three-dimensional models of \cite{fm}. It is the
statistics-changing $U_S(1)$-interaction, exclusive for
three-dimensional systems, which is strong enough to generate such
gaps. Unfortunately the $U_S(1)$ interaction does not have an
analogue in four-dimensions. 
Thus, although probably relevant for
the superconductivity scenaria, nevertheless such statistical
interactions may be not directly  relevant for the
four-dimensional physics underlying the findings of
\cite{krishana}. 

For our purposes below, therefore, we shall
ignore such non-Abelian statistical interactions, and concentrate
rather exclusively on the r\^ole of real electromagnetic
interactions, described by a potential $A_\mu$, in inducing, under
the influence of a strong external magnetic field, a mass gap for
the nodal holons. We shall assume, however, that the
electromagnetic interactions are screened along the $z$-direction
(interplanar), and in this sense we shall treat the gauge coupling
$e$ as spatially anisotropic.
A subtlety of this model is the relativistic non-invariance
of the electromagnetic interactions
in the presence of holons. The fermion (holon or, even, charged
electron excitations) part
is a relativistic system describing nodal excitations,
but for such a system the r\^ole of the limiting speed of `light' is played
by the fermi velocity $v_F$ of the nodes.
On the other hand, the real electromagnetic field
propagates with the velocity of light $c \gg v_F$
(for realistic systems $c \sim 10^4 v_F$, but for us $v_F$ will be
considered as a
phenomenological parameter).
When the combined system is expressed in terms of the
fermi velocity, the electromagnetic interaction
will be non relativistic, with the velocity of light $c$ appearing explicitly
in the Maxwell action. This leads to
a non-relativistic form for the photon propagator in the environment
of holons or equivalently to a non-relativistic form for the fermion
propagator if we rescale the fields.

Our approach in the present article
will be field theoretic, and we shall not attempt to make
further contact with the condensed-matter systems, apart from
the very generic features mentioned above.
However, as we shall see, our findings are interesting enough,
and indeed may be of use in attempts to explain
the phenomenon of \cite{krishana} by means of
the magnetic catalysis mechanism of relativistic
fermions~\cite{farakos,fkm,others}.
A brief discussion on such `phenomenology' will appear at the end
of our article.

At this stage the reader's attention is drawn to 
a very interesting recent work~\cite{mir2}, 
which analyses the phenomenon of chiral symmetry breaking 
on a brane domain wall, embedded in a higher-dimensional space time, 
in the absence of any external fields.
This situation is different from the one discussed here,
not only because there is no external field, but also 
because in the work of \cite{mir2} the fermions are 
completely localized on the brane, 
while the gauge fields propagate in the bulk. In contrast,  
in our case we allow interlayer hopping for fermions,
which is anisotropic, and also we have anisotropic gauge couplings.
In the scenaria of \cite{mir2}, 
the fermions could represent 
purely three-dimensional holons
of planar high-temperature cuprates, 
localised on the Cu-O planes.
In such scenaria, therefore, the spin-charge separation
would be an exclusive feature of the planes, not 
extended (even in the strongly anisotropic sense)
to four dimensions, but the statistical interactions
(represented by the gauge fields) could extend to
interplanar coupling situations.  
This would make an important physical difference from 
the scenaria discussed in the present article. 

The structure of the article is as follows: in section 2 we study
the magnetic catalysis phenomenon in relativistic gauge-fermions
systems in the presence of anisotropic four-dimensional gauge
couplings. In section 3 we demonstrate that, in the strongly
anisotropic case, there is a significant enhancement of the induced mass
gap $m_{dyn}$ (on the plane), as compared with the isotropic case.
In section 4 we study the system at finite temperature, and
compute the critical temperature $T_c$ above which the
magnetically-induced mass gap disappears. We  demonstrate that
$T_c \simeq m_{dyn}$, as expected on natural grounds. The
square-root scaling with the magnetic field intensity is also
demonstrated. In section 5 we attempt to make contact with
realistic condensed matter systems, and the experiments of
\cite{krishana}, by repeating the above analysis but for a
non-relativistic fermion system, coupled to a relativistic
electromagnetic field, again with anisotropic couplings, whose
quantum fluctuations are taken into account in the absence of any
other interactions. The r\^ole of the limiting speed of light 
for the fermions (holons or electrically-charged nodal excitations) is
played by the fermi velocity $v_F$, while, the speed of light is $c > v_F$. 
In this case the ratio $v_F/c <1$ is a phenomenological parameter
of the model. The main
features of the previous fully-relativistic case are maintained,
especially as far as the enhancement of the fermion gap is
concerned in the strongly anisotropic case. However, there are 
extra suppression factors by powers of $v_F/c$, relative to the 
relativistic case of section 2.  
Conclusions and some
discussion, with relevance to `phenomenology' of high-temperature
superconductivity, are presented in section 6. Some formal aspects
of the Schwinger-Dyson analysis for the anisotropic case are
presented in an Appendix.

\section{Anisotropic gap equation}

The Lagrangian density which includes
the anisotropy is, in the absence of an external field

\be\label{lagran}
{\cal L}=-\frac{1}{4}F_{\mu\nu}F^{\mu\nu}+\ol\psi\left[i\br\partial
-e\br A-x\left(i\partial_3-eA_3\right)\gamma^3-m\right]\psi,
\ee

\nin where $\br\partial = \gamma^\mu \partial_\mu$, 
$\mu,\nu =0, 1, \dots, 3$, $\gamma^\mu$ are four-dimensional 
$4\times 4$ Dirac matrices, $\psi $ are four-component spinors, 
and 
the parameter $x$ controls the anisotropy. The case $x=0$ corresponds
to a totally isotropic situation (the usual 
quantum electrodynamics (QED)),
while $x=1$ corresponds 
to a totally anisotropic one: the gauge field lives in 3+1
dimensions whereas the fermions are 
{\it effectively} localized in 2+1 dimensions.
We choose the $\gamma$ matrices $\gamma^i, i=0,1,2$ diagonal and the
matrix $\gamma^3$ non-diagonal, as in \cite{fkm}. In the totally anisotropic
case $x=1$, $\gamma^3$ does not appear in the action and the other
$\gamma$ matrices decompose
in two sets of two-component Dirac matrices, appropriate
for the irreducible Dirac algebra in $(2+1)$-dimensions. 
In such a case the four-component spinors also decompose to 
an even number (2 for a single flavour) of two-component 
(2+1)-dimensional spinors, and one recovers the 
planar case, after integration over the third dimension~\footnote{In the 
physical case of relevance to condensed matter, 
the third dimension may extend between two Cu-O layers, which 
play the r\^ole of boundaries of the available space. In such a case
the extra (third) dimension is integrated in the bulk space between these boundaries. In the analysis below we shall not assume 
explicitly such geometries, but
rather treat the anisotropic four-dimensional case in a generic sense,
specified by the Lagrangian (\ref{lagran}).}.

It should be stressed that Eq.(\ref{lagran}) of course respects the gauge invariance. With such a
Lagrangian, and in the absence of external field, the bare fermion
propagator is given by

\be\label{propbare}
iS^{-1}(p)=\br p-xp_3\gamma^3-m.
\ee

\nin As an illustration of the effect of the anisotropy, we can
compute the propagation of a fermion in the direction 3. Let us
define in the Euclidean space
\bea
\phi(r_3)&=&\mbox{tr}\int dr_4dr_2dr_1 <0|\ol\psi(0)\psi(r)|0>\nonu
&=&\mbox{tr}\int dr_4dr_2dr_1\int\frac{d^4p}{(2\pi)^4} S(p)e^{ipr}\nonu
&=&\mbox{tr}\int\frac{dp_3}{2\pi}S(0,0,p_3,0)e^{ip_3r_3},
\eea

\nin i.e. the propagation rate of the fermions from the 2+1 dimensional
worlds $r_3=0$ and $r_3\ne 0$. A straightforward computation leads to

\be
\phi(r_3)=\frac{2}{1-x}\exp\left(-\frac{m|r_3|}{1-x}\right),
\ee

\nin which shows that as the anisotropy increases, the fermion
propagation in the direction 3 decreases exponentially.
Eventually, as $x\to 1$, this propagation vanishes if $r_3\ne 0$.
This corresponds to an effective delta-function in $r_3$ 
which kills the
$r_3$ integration in the action leading to a three-dimensional theory. 

We derive in the appendix the Schwinger-Dyson equation for the
fermion propagator, taking into account the anisotropy parameter $x$
and find (we do not write the space-time indices)

\be\label{SD}
G=S-4\pi\alpha\int S\gamma^\mu G\Lambda^\nu GD_{\mu\nu}
+x4\pi\alpha\int S\gamma^3 G\Lambda^\nu GD_{3\nu}
\ee

\nin where $G$ is the full fermion propagator,
$D_{\mu\nu}$ the full photon propagator and $\Lambda^\mu$ the full
vertex which satisfies at the tree level

\bea
\Lambda^\mu_{tree}&=&\gamma^\mu~~~~\mbox{if}~~\mu\ne 3\nonu
\Lambda^3_{tree}&=&(1-x)\gamma^3
\eea

\nin We use the
usual definition of the fine structure constant: $e^2=4\pi\alpha$.

In the presence of an external, constant
and homogenous magnetic field in the direction 3, we can choose an external
gauge such that $A^{ext}_3=0$ for which the parameter $x$ will not couple
to the external field and thus will only play a role in the substitution
$p_3\to(1-x)p_3$. The lowest Landau level (LLL)
approximation \cite{miransky} for the
fermion propagator will be then

\be
S^L(y,z)=e^{iey^\mu A^{ext}_\mu(z)}\tilde S^L(y-z),
\ee

\nin where the Fourier transform of the translational invariant
propagator $\tilde S^L$ is

\be\label{LLL}
\tilde S^L(p)=e^{-p_\bot^2/|eB|}
\frac{i(1-i\gamma^1\gamma^2)}
{p_0\gamma^0+(1-x)p_3\gamma^3-m}
\ee

\nin where $p_\bot^2=p_1^2+p_2^2$ is the transverse momentum squared. 
We will take $m=0$ for the bare propagator (not taking into
account the interaction with the dynamical gauge field) and $m=m(p_0,p_3)$,
the dynamical self energy which depends only on the
longitudinal momenta in the LLL approximation, for the full propagator.
The integral equation describing the magnetic catalysis in the LLL
approximation contains only the components $D_{00}$ and $D_{33}$ of the
photon propagator \cite{miransky},
due to the spin projector $(1-i\gamma^1\gamma^2)/2$
in the fermion propagator (\ref{LLL}). These two components
lead to equal contributions and thus the second integral in the
Schwinger-Dyson equation (\ref{SD}) which is proportional to $x$ gives
half the contribution of the first integral. The integral equation
reads then, if we neglect the corrections to the vertex,

\be\label{sdLLL}
G^L=S^L-4\pi\alpha\left(1-x+\frac{x^2}{2}\right)
\int S^L\gamma^\|G^L\gamma^\|G^LD_{\|,\|},
\ee

\nin where $\|$ denotes the longitudinal components. The final
integral equation is found by making the substitution $p_3\to
(1-x)p_3$ in the fermion propagators appearing in the integral of
Eq.(\ref{sdLLL}) which has already been computed in the isotropic
case \cite{gusynin}. If we rescale all the quantities homogenous
to a mass by $\sqrt{|eB|}$, we have in the Euclidean space

\bea
\mu(k_3,k_4)&=&\left(1-x+\frac{x^2}{2}\right)\frac{\alpha}{\pi^2}
\int \frac{dp_3dp_4 \mu(p_3,p_4)}{(1-x)^2p_3^2+p_4^2+\mu^2(p_3,p_4)}\nonu
&&\times\int_0^\infty \frac{du e^{-u}}{2u+(p_3-k_3)^2+(p_4-k_4)^2},
\eea

\nin where $\mu$ is the dimensionless self-energy and $u$ a dimensionless
transverse momentum squared.
The momenta $p_3$ and $p_4$ play
a symmetric role in the photon propagator whereas they
enter non-symmetrically in the fermion propagator, which leads to
a quite difficult analysis of the integral equation if we wish to
take into account the momentum dependence of the fermion self energy. We
know that this momentum dependence is essential in 3+1 dimensions
\cite{miransky} but not in 2+1 dimensions, as long as we consider a
qualitative description \cite{afk3}. In this paper we wish to give a
qualitative description of the magnetic catalysis in the strongly
anisotropic regime $1-x<<1$ and thus will make the constant self-energy
approximation in which the integral equation finally reads

\be\label{intequa}
1=\left(1-x+\frac{x^2}{2}\right)\frac{\alpha}{\pi^2}
\int \frac{dp_3dp_4}{(1-x)^2p_3^2+p_4^2+\mu^2_0}
\int_0^\infty \frac{du e^{-u}}{2u+p_3^2+p_4^2},
\ee

\nin or, when we make the angular integration,

\be\label{intequaang}
1=\left(1-x+\frac{x^2}{2}\right)\frac{\alpha}{\pi}\int_0^\infty
\frac{d\rho}{\sqrt{(\rho+\mu_0^2)[(1-x)^2\rho+\mu_0^2]}}
\int_0^\infty \frac{du e^{-u}}{2u+\rho}
\ee

\nin The analysis of the approximation (\ref{intequa})
in the isotropic regime $x<<1$ will not
lead to a reliable quantitative dynamical mass $\mu_0$ but
will show us the qualitative tendency
of $\mu_0$ to increase when the anisotropy increases.

\section{Enhancement of the dynamical mass}

In this section we will show that the anisotropy generates a 
considerably enhanced 
mass gap, compared to the isotropic situation where we know that in the
constant self-energy approximation \cite{gusynin}, \cite{lee},

\be
\mu_0(x=0)\simeq \sqrt 2\exp\left(-\sqrt\frac{\pi}{\alpha}\right).
\ee

\nin To see the increase of the dynamical mass when $x$ increases,
we make an expansion of the integral equation (\ref{intequa}) up
to the order $x^2$ and find

\be
1=\frac{\alpha}{\pi^2}\int_0^\infty du e^{-u}\left\{{\cal I}_1
+x\left(2{\cal I}_2-{\cal I}_1\right)
+x^2\left(4{\cal I}_3-3{\cal I}_2+\frac{1}{2}{\cal I}_1\right)+...\right\},
\ee

\nin where

\bea\label{calI}
{\cal I}_1&=&\int\frac{dp_3dp_4}{(p_3^2+p_4^2+\mu_0^2)(p_3^2+p_4^2+2u)}\nonu
&=&\pi\frac{\ln(2u/\mu_0^2)}{2u-\mu_0^2}\nonu
{\cal I}_2&=&\int\frac{dp_3dp_4p_3^2}{(p_3^2+p_4^2+\mu_0^2)^2
(p_3^2+p_4^2+2u)}\nonu
&=&\frac{\pi}{2}\frac{2u\ln(2u/\mu_0^2)-2u+\mu_0^2}
{(2u-\mu_0^2)^2}\\
{\cal I}_3&=&\int\frac{dp_3dp_4p_3^4}{(p_3^2+p_4^2+\mu_0^2)^3
(p_3^2+p_4^2+2u)}\nonu
&=&\frac{3\pi}{16}\frac{8u^2\ln(2u/\mu_0^2)
+(\mu_0^2-6u)(2u-\mu_0^2)}{(2u-\mu_0^2)^3}.\nonumber
\eea

\nin Despite their appearance, the integrals (\ref{calI}) are
converging when $2u\to\mu_0^2$, what can be checked by expanding
the logarithms around $2u=\mu_0^2$.

For the integration over the transverse momentum $u$, we write
that for $\mu_0<<1$:

\bea
&&\int_0^\infty du {\cal I}_1 e^{-u}
=\pi\int_0^\infty du e^{-u}\frac{\ln(2u/\mu_0^2)}{2u-\mu_0^2}\nonu
&&\simeq \frac{\pi}{2}\int_0^{2/\mu_0^2}du\frac{\ln u}{u-1}
\simeq \frac{\pi}{2}\int_1^{2/\mu_0^2}du\frac{\ln u}{u}\nonu
&&=\pi\ln^2\left(\frac{\sqrt 2}{\mu_0}\right).
\eea

\nin Similar approximations lead to

\bea
\int_0^\infty du {\cal I}_2 e^{-u}&\simeq&\frac{\pi}{2}
\ln^2\left(\frac{\sqrt 2}{\mu_0}\right)\nonu
\int_0^\infty du {\cal I}_3 e^{-u}&\simeq&\frac{3\pi}{8}
\ln^2\left(\frac{\sqrt 2}{\mu_0}\right),
\eea

\nin and give the following equation for the dynamical mass

\be\label{equafinale}
\left(1+\frac{x^2}{2}+...\right)\ln^2\left(\frac{\sqrt 2}{\mu_0}\right)
=\frac{\pi}{\alpha}.
\ee

\nin We have then for the dimensionful dynamical mass, to order $x^2$,

\be\label{resfinal}
m_{dyn}(x)\simeq\sqrt 2\sqrt{|eB|}\exp\left\{-\sqrt\frac{\pi}{\alpha}
\left(1-\frac{x^2}{4}\right)\right\}.
\ee

\nin We see then that the
anisotropy ($x>0$) has the effect to
increase the dynamical mass.

Let us now come to the totally anisotropic regime and compute the
dynamical mass obtained for $x=1$.
In this case, we find from (\ref{intequaang})

\be\label{anisointequa}
1=\frac{\alpha}{2\pi}\int_0^\infty\frac{d\rho}{\sqrt{\rho+1}}
\int_0^\infty\frac{du e^{-u}}{2u+\rho\mu_0^2}.
\ee

\nin To perform the integration over $\rho$, we suppose that $2u>\mu_0^2$,
which consists in neglecting the interval $[0,\mu_0^2/2]$ in the integration
over $u$, which will be justified since we will find $\mu_0^2<<1$.
We have then

\be
1\simeq\frac{\alpha}{2\mu_0}\int_{\mu_0^2/2}^\infty du e^{-u}
\frac{1}{\sqrt{2u-\mu_0^2}}\simeq\frac{\alpha}{4}\int_1^{2/\mu_0^2}
du \frac{1}{\sqrt{u-1}},
\ee

\nin such that we finally obtain for the dimensionful dynamical mass

\be\label{anisodynmass}
m_{dyn}(x=1)\simeq\frac{\alpha}{\sqrt 2}\sqrt{|eB|}.
\ee

\begin{figure}
\epsfxsize=10cm
\epsfysize=8cm
\centerline{\epsfbox{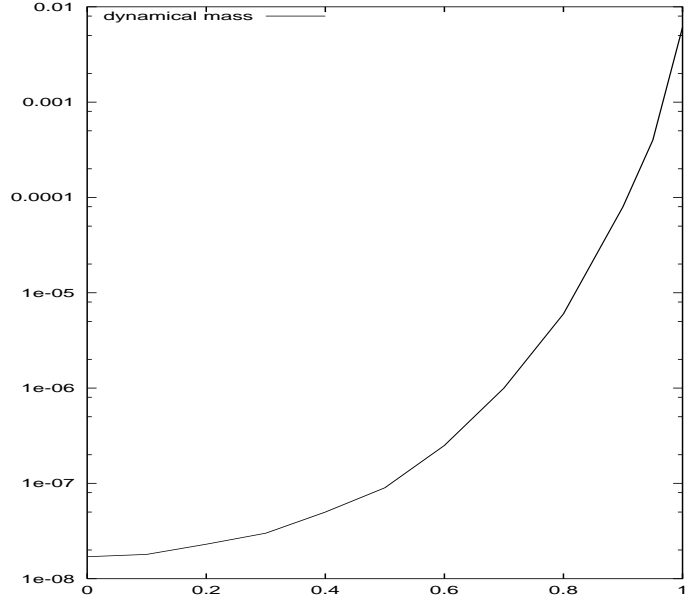}}
\caption{$\mu_0$ versus $x$ for $\alpha=.01$}
\label{layer}
\end{figure}

\nin This result shows that the anisotropy plays a fundamental role in the
generation of a big mass gap since $m_{dyn}(x=1)>>m_{dyn}(x=0)$. We show in
figure \ref{layer} the dynamical mass as a function of $x$, for a given
value of the coupling $\alpha$. This figure has been done with the
numerical study of Eq.(\ref{intequaang}) and we note the logarithmic
scale, showing the exponential increase of the dynamical mass. The
analytical results (\ref{resfinal}) and (\ref{anisodynmass})
are confirmed by the curve
in the asymptotic limits $x<<1$ and $1-x<<1$.

\section{Critical temperature in the anisotropic regime}

We now turn to the finite temperature treatment of the problem so as to
find the critical temperature when $x=1$.

The integral equation at finite temperature is obtained by the
usual substitutions $p_4\to\omega_l=(2l+1)\pi t$ and
$\int dp_4\to 2\pi t\sum_l$ in the equation (\ref{intequa}), and
$t=T/\sqrt{|eB|}$ is the dimensionless temperature. We just have to
pay attention to the photonic momentum $p_4$ which has to be
changed by $\omega_l-\omega_0$, that is to say to a bosonic
Matsubara mode.
We obtain then the following equation for the critical temperature
defined by $\mu_0(t_c)=0$:

\be\label{finiteT}
1=\frac{\alpha}{\pi}\int dp_3\int_0^\infty du e^{-u}
t_c\sum_{l=-\infty}^{l=\infty}
\frac{1}{\omega_l^2[2u+p_3^2+(\omega_l-\omega_0)^2]}
\ee

\nin where $\omega_0=\pi t_c$.
The summation over the dimensionless Matsubara modes $\omega_l$
is done with the usual contour deformation trick detailed for
example in \cite{afk3} and gives

\bea
&&t_c\sum_{l=-\infty}^{l=\infty}
\frac{1}{\omega_l^2[2u+p_3^2+(\omega_l-\omega_0)^2]}=
\frac{1}{4t_c}\frac{1}{2u+p_3^2+\omega_0^2}\\
&&+\frac{1}{2\sqrt{2u+p_3^2}}\coth\left(\frac{\sqrt{2u+p_3^2}}{2t_c}\right)
\frac{\omega_0^2-2u-p_3^2}{(\omega_0^2+2u+p_3^2)^2}.\nonumber
\eea

\nin We suppose that the $\coth$ is close to 1 since $t_c<<1$.
This is actually not valid when $2u+p_3^2<<1$ but this region is
negligible in the remaining integrals. The integral equation reads
then

\bea\label{intert}
1&=&\frac{\alpha\pi}{8}\int_0^\infty du e^{-u\omega_0^2/2}
\frac{1}{\sqrt{1+u}}\nonu
&&+\frac{\alpha}{\pi}\int_0^\infty du e^{-u\omega_0^2/2}
\int_0^\infty dv \frac{1-u-v^2}{(1+u+v^2)^2\sqrt{u+v^2}}
\eea

\nin The dominant contribution in Eq.(\ref{intert}) comes from
the first term when $t_c<<1$ since in the second term, the
integration over $v$ gives

\bea
&&\int_0^\infty dv \frac{1-u-v^2}{(1+u+v^2)^2\sqrt{u+v^2}}\nonu
=&&-\frac{1}{1+u}+\frac{1}{4(1+u)^{3/2}}\ln\left(\frac{2+u+2\sqrt{1+u}}
{2+u-2\sqrt{1+u}}\right),
\eea

\nin which leads to higher powers of $t_c$.
The critical temperature is then given by

\be
1\simeq\frac{\alpha\pi}{8}\int_0^{2/\omega_0^2}du\frac{1}{\sqrt{1+u}},
\ee

\nin such that the dimensionful critical temperature is finally

\be\label{tc}
T_c(x=1)\simeq\frac{\alpha}{2\sqrt 2}\sqrt{|eB|}=\hf m_{dyn}(x=1).
\ee

\nin This result confirms what was expected, i.e. the critical temperature
is of the order of the dynamical mass at zero temperature.
The same conclusion was found in \cite{gus} for isotropic QED.

\section{Non-Relativistic Gauge Field--Fermion \\ Models and an Application to
Condensed Matter}

So far we have examined a relativistic system,
in which the gauge interactions may be assumed distinct
from the real electromagnetic interactions.
From a condensed matter view point, such systems may
have some relevance to effective gauge theories of $t-j$ models,
with intrasublattice hopping, pertaining to spin-charge separation
scenaria
applied, however, to nodal $d$-wave excitations~\cite{wiegmann}.

In this section we shall consider a problem which might be
considered as more ``realistic'', in the sense of being connected
directly to the observed physics of high-temperature
superconductors, in view of the recent experimental findings of
\cite{krishana}. We shall not assume {\it ad hoc} any statistical
interaction, but we shall continue to apply  a spin-charge
separation scenario~\cite{anderson} for the nodal excitations of
the high-temperature $d$-wave superconductors.

In this case, the electrically-charged
excitations (holons)
around the nodes of a $d$-wave supercondcuting gap will be
represented by $(3+1)$-dimensional Dirac fermions coupled to
an electromagnetic field, described by a potential
$A_\mu = (A_0, A_i)$.  
The four-dimensional nature of the fermions implies the possibility
of interplanar hopping for such 
excitations in the materials, which is a realistic feature. 
We shall consider the case where
the electromagnetic field has a background, corresponding to a constant
magnetic field along the $z$ direction, and four-dimensional
quantum fluctuations around it, and assume no further interactions
among the holons.
However,
the electromagnetic coupling
$e$ will be assumed screened along the $z$-direction, i.e. the coupling
is anisotropic.
Such a screening may be provided by a combination
of both the chemistry and geometry of the
planar materials, which consist of layers of Cu-O,
with lattice spacing of $5$ Angstr\"oms, while the interlayer distance
is much larger, of the order of $100$ Angstr\"oms.
In the interlayer space there are doping atoms {\it etc}.,
which may be responsible
for an effective charge screening.

From a formal view point we might think of the effective
electric-charge anisotropy as follows: the coupling constant
$e^2/4\pi$ entering the problem, as a result of quantum
fluctuations of the electromagnetic field, is actually a running
coupling constant, which depends on the available energy. If the
latter is viewed as the inverse of a characteristic distance for
the problem, we then observe that, as a result of the relatively
large separation between two holons (which are the relevant
degrees of freedom for the problem) at different layers, assumed
perpendicular to the $z$ direction, the resulting effective charge
(along the $z$ direction) lies in the infrared regime, and as such
is much weaker than the effective charge between holons in the
same layer and actually at characteristic distances of the order
of the inverse of the magnetically-induced mass gap, which we are
interested in for the purposes of the present work. In
high-temperature  materials this distance, i.e. the magnetic
coherence length of nodal charged excitations, is short enough and
of the order of a few angstr\"oms, which justifies the assumed
strong anisotropy of the electric charge.

With these in mind, we consider the following model.
The photon kinetic term is the usual one

\be
{\cal L}_G=-\frac{1}{4}F_{\mu\nu}F^{\mu\nu},
\ee

\nin with $\partial_0=\partial/\partial t$ ($c=1$) and the
free fermion kinetic term is

\be
{\cal L}_F=\ol\psi\left(i\gamma^0\tilde\partial_0+
i\gamma^k\partial_k\right)\psi
\ee

\nin with $\tilde\partial_0=\partial/\partial(v_Ft)$.
Since the fermion density and current are
(the notation $\vec{A} $ denotes spatial three vectors): 
\bea
\rho&=&\psi\dg\psi\nonu
\vec {\j}&=&v_F\ol\psi\vec\gamma\psi,
\eea

\nin the interaction between the fermions and the gauge field is
\be
{\cal L}_I=e\ol\psi\gamma^0 A_0\psi+ ev_F\ol\psi\vec\gamma.\vec A\psi
\ee

\nin such that the final Lagrangian will contain the following
fermionic part

\be
\ol\psi\left[\eta\left(i\partial_0\gamma^0-g A_0\gamma^0\right)
+\left(i\partial_k\gamma^k-g A_k\gamma^k\right)
-x\left(i\partial_3\gamma^3-g A_3\gamma^3\right)\right]\psi,
\ee

\nin where we define $\eta=1/v_F$, $g=ev_F$ and $k=1,2,3$.
The bare fermion propagator is then, in the absence of external field,

\be
iS^{-1}(p)=\eta p_0\gamma^0+p_k\gamma^k-xp_3\gamma^3-m
\ee
It is straightforward to see from
the derivation shown in the appendix that the
Schwinger-Dyson equation changes to

\bea
G&=&S-\eta 4\pi\alpha\int S\gamma^0G\Lambda^\nu G D_{0\nu}\nonu
&&-4\pi\alpha\int S\gamma^kG\Lambda^\nu G D_{k\nu}
+x 4\pi\alpha\int S\gamma^3G\Lambda^\nu D_{3\nu},
\eea

\nin where

\bea
\Lambda^\nu_{tree}&=&\gamma^\nu~~~~\mbox{if}~~\nu\ne 0,3\nonu
\Lambda^0_{tree}&=&\eta\gamma^0\nonu
\Lambda^3_{tree}&=&(1-x)\gamma^3,
\eea

\nin and $g^2=4\pi\alpha$.
The integral equation corresponding to (\ref{intequa})
is finally

\be\label{etaintequa}
1=[\eta^2+(1-x)^2]\frac{\alpha}{2\pi^2}\int
\frac{dp_3dp_4}{(1-x)^2p_3^2+\eta^2p_4^2+\mu_0^2}\int_0^\infty
\frac{du e^{-u}}{2u+p_3^2+p_4^2}.
\ee

\nin This last equation reads for $x=1$

\be
1=\frac{\alpha}{2\pi^2}\int\frac{dp_3dp_4}{p_4^2+(\mu_0/\eta)^2}
\int_0^\infty \frac{du e^{-u}}{2u+p_3^2+p_4^2},
\ee

\nin which is the equation (\ref{anisointequa}) with the
substitution $\mu_0\to\mu_0/\eta$.
Since $\eta>>1$, the condition $\mu_0/\eta<<1$ is still valid and
the result is then for the dimensionful dynamical mass

\be\label{mdyn} 
m_{dyn}\simeq v_F^{3/2}\frac{e^2}{4\pi\sqrt 2}\sqrt{|eB|}.
\ee

For the finite temperature case, the equation (\ref{finiteT}) does not
change (the factors $\eta$ cancel when $x=1$)
and the critical temperature is then

\be\label{tcrit}
T_c\simeq v_F^{5/2}\frac{e^2}{8\pi\sqrt 2}\sqrt{|eB|}
\ee

It is important to notice  that both the 
the dynamical mass (\ref{mdyn}), and the associated
critical temperature (\ref{tcrit}), are proportional 
to a suppression factor $v_F^{3/2}={\cal O} (10^{-6})$ and
$v_F^{5/2} ={\cal O}(10^{-10})$ respectively, compared
to the previoulsy considered relativistic case. 
This may have important phenomenological implications when one 
attempts to compare the scenaria advocated here with realistic
condensed-matter situations~\cite{krishana}. 
We shall not do such analyses here, given that 
at present we lack a detailed derivation of the 
continuum models discussed above from microscopic condensed
matter models. This is essential for providing the 
correct order of magnitude of the various coupling constants,
such as gauge, hopping elements {\it etc.}, entering the model. 

\section{Discussion}

In this work we have discussed the case of four-dimensional 
anisotropic fermions coupled to anisotropic-coupling 
Abelian gauge fields, and external 
magnetic fields. By considering fluctuations of the gauge fields
we have considered the magnetic catalysis phenomenon, i.e. the 
dynamical mass generation for fermions
under the influence of strong external magnetic fields. 
We have considered two cases: 
the fully relativistic case, in which fermions and fluctuating
gauge fields are relativistic, and the case
where the photon fields are relativistic, but the fermion 
part of the Lagrangian is non relativistic, with a 
fermi velocity $v_F \ne c$
($c$ is the speed of light) playing the r\^ole of the 
limiting velocity for the fermionic part. 

In both cases a strong planar anisotropy
has been assumed. The analysis has shown 
a significant enhancement of the induced mass gap, relative 
to the isotropic four-dimensional case. In the non-relativistic 
case, however we find that 
the induced mass gap is found suppressed by some power 
of the fermi velocity $v_F < c$, as compared to the 
relativistic case. In realistic situations, $v_F/c \sim 10^{-4}$,
and the suppression factors are significant.
In both cases, the mass gap and the associated critical
temperature, above which the magnetic catalysis 
disappears due to thermal disorder, exhibit a square-root
scaling with the magnetic field intensity, as expected  
on natural grounds. 

The non relativistic case, 
may be related to the situation encountered in 
high-temperature superconductors in
the experiments of \cite{krishana}. 
In such a case, the fermions are electrically
charged excitations, which can represent 
excitations about the nodes of the $d$-wave superconducting 
gaps. The relativistic nature of the excitations is due to the 
nodal structure. These excitations may be 
holons, capable though of interlayer hopping, 
or could be real (nodal) electron degrees of freedom. 
In this case the fluctuating gauge fields are 
assumed electromagnetic in origin.

The fully relativistic case, examined in section 2, 
may admit a less conventional physical interpretation, and 
correspond to the case considered in 
\cite{wiegmann}. There, the abelian gauge interactions 
represent magnetic interactions in a spin-charge
separation framework for high-temperature superconductors,
in which one admits intrasublattice hopping in the 
underlying doped antiferromagnetic model. 
In contrast to that work, however, which used 
non-relativistic fermions, 
here  the fermions (holons)
are relativistic, since one considers the spin-charge separation
near the nodes. One also assumes a four-dimensional nature for such fermions,
as a result of interplanar hopping. 
However, the above interpretation of the relativistic
case is still not fully understood in the sense of
not having been derived by an appropriate microscopic model
at a satisfactory level of mathematical rigor.

At any rate, the generic results obtained 
from the field-theoretic analysis of the present article 
may be useful if one 
wishes to compare them with 
various scenaria that may be in operation
in realistic condensed-matter 
situations, namely detailed 
phenomenological models derived from microscopic 
condensed-matter systems with relevance to doped antiferromagnets
(and thus high-temperature superconductivity).
In such cases it would be also useful to examine models where 
one has non-Abelian ($SU(2)$ type) four-dimensional  
fluctuating gauge fields, coupled to fermions, in the presence
of external magnetic fields. This would be more relevant 
to particle-hole symmetric spin-charge separating doped 
models, such as those considered in~\cite{farakos}. 
We hope to come to a detailed discussion of such issues in a
forthcoming publication.

\par\noindent
\section*{Acknowledgements}\par\noindent

The work of J.A.,  K.F. and G.K. is partially supported
by the TMR project FMRX-CT97-0122.
That of N.E.M. is partially supported by
the Leverhulme Trust (U.K.). K.F. wishes to thank
Sarben Sarkar and the Physics Department of King's College London
for the hospitality during the initial stages of this work.

\par\noindent

\begin{appendix}

\section{Anisotropic Schwinger-Dyson equation}

In this Appendix we discuss 
the derivation of the Schwinger-Dyson equation for the fermion propagator in
the anisotropic case. To this end, we
shall follow the derivation given in \cite{zuber} for the isotropic case.
Let us first recall some
definitions.
Starting from the Lagrangian

\be
{\cal L}=-\frac{1}{4}{\cal F}_{\mu\nu}{\cal F}^{\mu\nu}
+\ol\Psi\left[i\br\partial
-g\br {\cal A}-x\left(i\partial_3-g{\cal A}_3\right)\gamma^3-m\right]\Psi,
\ee

\nin we define the connected graphs generator functional $W$ by

\be\label{cggen}
\exp W[\ol\eta,\eta,j_\mu]
=\int{\cal D}[{\cal A}_\mu,\ol\Psi,\Psi]
\exp\left\{i\int_x{\cal L}+i\int_x(j^\mu{\cal A}_\mu+\ol\eta\Psi+
\ol\Psi\eta)\right\}\nonumber.
\ee

\nin $W$ has the following functional derivatives  (we do not
write the space-time indices)

\bea\label{derivw}
\frac{\delta W}{\delta j^\mu}&=&\frac{1}{Z}\left<i{\cal A}_\mu\right>
=iA_\mu\nonu
\frac{\delta W}{\delta\ol\eta}&=&\frac{1}{Z}\left<i\Psi\right>=i\psi\nonu
W\frac{\la\delta}{\delta\eta}&=&\frac{1}{Z}\left<i\ol\Psi\right>=i\ol\psi\nonu
\frac{\delta}{\delta\ol\eta}W\frac{\la\delta}{\delta\eta}&=&-\ol\psi\psi
+\frac{1}{Z}\left<\ol\Psi\Psi\right>,
\eea

\nin where the expectation value $\left<{\cal O}\right>$ of an operator
${\cal O}$ is

\be
\left<{\cal O}\right>=\int{\cal D}[{\cal A},\ol\Psi,\Psi]
~{\cal O}~\exp\left\{i\int_x{\cal L}+
i\int_x(j{\cal A}+\ol\eta\Psi+\ol\Psi\eta)\right\},
\ee

\nin and we define

\be
(\ol\eta\eta)\frac{\la\delta}{\delta\eta}=
-\frac{\delta}{\delta\eta}(\ol\eta\eta)=\ol\eta.
\ee

\nin Inverting the relations between $(j_\mu,\ol\eta,\eta)$
and $(A_\mu,\ol\psi,\psi)$,
we define the effective action $\Gamma[A_\mu,\ol\psi,\psi]$ as the
Legendre transform
of $W[j_\mu,\ol\eta,\eta]$ by

\be
W=i\Gamma+i\int_x\left(j^\mu A_\mu+\ol\eta\psi+\ol\psi\eta\right).
\ee

\nin From this definition we extract the following functional derivatives:

\bea\label{derivg}
\frac{\delta\Gamma}{\delta A_\mu}&=&-j^\mu\nonu
\frac{\delta\Gamma}{\delta\ol\psi}&=&-\eta\nonu
\Gamma\frac{\la\delta}{\delta\psi}&=&-\ol\eta\nonu
\frac{\delta}{\delta\ol\psi}\Gamma\frac{\la\delta}{\delta\psi}&=&
-\frac{\delta\ol\eta}{\delta\ol\psi}
=-i\left(\frac{\delta}{\delta\ol\eta}W
\frac{\la\delta}{\delta\eta}\right)^{-1}.
\eea

\nin The starting point to derive the Schwinger-Dyson equation
is to assume that the integral of a derivative
vanishes, such that we can write

\be
\int{\cal D}[{\cal A}_\mu,\ol\Psi,\Psi]\frac{\delta}{\delta\ol\Psi(z)}
\exp\left\{i\int{\cal L}+i\int j^\mu {\cal A}_\mu
+\ol\Psi\eta+\ol\eta\Psi\right\}=0,
\ee

\nin which leads to, after a functional derivative with respect to
$\eta$ and setting $j^\mu=\eta=\ol\eta=0$,

\bea\label{sdint}
&&\delta(z_1-z_2)=\left(i\br\partial_{z_1}-m-ix\partial_3\gamma^3\right)
\frac{\delta}{\delta\ol\eta(z_1)}W\frac{\la\delta}{\delta\eta(z_2)}
|_{j^\mu=\eta=\ol\eta=0}\\
&&+ig\left(\frac{\delta}{i\delta j^\mu(z_1)}\gamma^\mu-x
\frac{\delta}{i\delta j^3(z_1)}\gamma^3\right)
\frac{\delta}{\delta\ol\eta(z_1)}W\frac{\la\delta}{\delta\eta(z_2)}
|_{j^\mu=\eta=\ol\eta=0}\nonumber.
\eea

\nin We now have to turn this last equation into an equation for
the effective action if we
wish to obtain a relation between the proper functions. For this, we
will use the last equation of Eq.(\ref{derivg}) and we remark
that

\bea
\frac{\delta \Gamma}{\delta j_\mu(z)}&=&
\int dy \left\{\frac{\delta\Gamma}{\delta A_\nu(y)}
\frac{\delta A_\nu(y)}{\delta j_\mu(z)}+
\frac{\delta\ol\psi(y)}{\delta j_\mu(z)}
\frac{\delta\Gamma}{\delta\ol\psi(y)}+\frac{\Gamma\la\delta}{\delta\psi(y)}
\frac{\delta\psi(y)}{\delta j_\mu(z)}\right\}\nonu
&=&\int dy \left\{
\frac{\delta\Gamma}{\delta A_\nu(y)}\left(\frac{\delta^2\Gamma}
{\delta A_\mu(z)\delta A_\nu(y)}\right)^{-1}\right.\nonu
&&~~~~~~~~+\left(\frac{\delta^2\Gamma}
{\delta A_\mu(z)\delta\ol\psi(y)}\right)^{-1}
\frac{\delta\Gamma}{\delta\ol\psi(y)}\nonu
&&~~~~~~~~\left.+\frac{\Gamma\la\delta}{\delta\psi(y)}
\left(\frac{\delta\Gamma\la\delta}
{\delta A_\mu(z)\delta\psi(y)}\right)^{-1}\right\},
\eea

\nin such that we have for vanishing sources

\bea
&&\frac{\delta^2\Gamma\la\delta}{\delta\ol\psi(z_1)\delta j_\mu(z_2)
\delta\psi(z_3)}|_{\psi=\ol\psi=A_\mu=0}\nonu
&&=\int dy
\frac{\delta^2\Gamma\la\delta}{\delta\ol\psi(z_1)\delta A_\nu(y)
\delta\psi(z_3)}\left(\frac{\delta^2\Gamma}{\delta A_\mu(z_2)
\delta A_\nu(y)}\right)^{-1}.
\eea

\nin The proper functions (respectively fermion propagator, photon
propagator and vertex) are defined by

\bea
G^{-1}(z_1,z_2)&=&-i\frac{\delta}{\delta\ol\psi(z_1)}\Gamma\frac{\la\delta}
{\delta\psi(z_2)}|_{\ol\psi=\psi=A_\mu=0}\\
D^{-1}_{\mu\nu}(z_1,z_2)&=&-i\frac{\delta^2\Gamma}{\delta A^\mu(z_1)
\delta A^\nu(z_2)}|_{\ol\psi=\psi=A_\mu=0}\nonu
\Lambda^\mu(z_1;z_2,z_3)&=&-\frac{1}{g}\frac{\delta^2\Gamma\la\delta}
{\delta A_\mu(z_1)\delta\ol\psi(z_2)
\delta\psi(z_3)}|_{\ol\psi=\psi=A_\mu=0}\nonumber
\eea

\nin such that we finally obtain from Eq.(\ref{sdint})
after multiplying by $S$ (we do not write the space-time indices)

\be
G=S-4\pi\alpha\int S\gamma^\mu G\Lambda^\nu G D_{\mu\nu}
+x4\pi\alpha\int S\gamma^3 G\Lambda^\nu G D_{3\nu}
\ee

\nin where $g^2=4\pi\alpha$.

\end{appendix}

\end{document}